\documentclass[prl,twocolumn,showpacs,superscriptaddress]{revtex4}
\usepackage[dvips]{epsfig}
\usepackage{graphicx}
\usepackage{amsfonts}
\usepackage{bm}
\usepackage{amsmath,amssymb,lscape,float}

\newcommand{\dg}{\dagger}

\makeatletter
\renewcommand\appendix{\par
  \setcounter{section}{0}  \setcounter{subsection}{0}  \setcounter{equation}{0}  \gdef\thesection{\Alph{section}}
  \@addtoreset {equation}{section}
  \renewcommand{\theequation}{\thesection\arabic{equation}}}
\makeatother

\begin{document}
\title{Quantum simulation of small-polaron formation with trapped ions}
\author{Vladimir M. Stojanovi\'c}
\email{vladimir.stojanovic@unibas.ch}
\affiliation{Department of Physics, University of Basel, Klingelbergstrasse 82, CH-4056
Basel, Switzerland}
\author{Tao Shi}
\email{tao.shi@mpq.mpg.de}
\affiliation{Max-Planck-Institut f\"{u}r Quantenoptik, Hans-Kopfermann-Str. 1, 85748
Garching, Germany}
\author{C. Bruder}
\affiliation{Department of Physics, University of Basel, Klingelbergstrasse 82, CH-4056
Basel, Switzerland}
\author{J. Ignacio Cirac}
\affiliation{Max-Planck-Institut f\"{u}r Quantenoptik, Hans-Kopfermann-Str. 1, 85748
Garching, Germany}
\date{\today }

\begin{abstract}
We propose a quantum simulation of small-polaron physics using a
one-dimensional system of trapped ions acted upon by off-resonant standing
waves. This system, envisioned as an array of microtraps, in the
single-excitation case allows the realization of the anti-adiabatic regime
of the Holstein model. We show that the strong excitation-phonon coupling
regime, characterized by the formation of small polarons, can be reached
using realistic values of the relevant system parameters. Finally,
we propose measurements of the quasiparticle residue and the average 
number of phonons in the ground state, experimental probes validating the 
polaronic character of the phonon-dressed excitation.
\end{abstract}

\maketitle

The field of quantum simulation~\cite{Lloyd:96}, inspired by the ideas of
Feynman~\cite{feynmansim}, holds promise to advance our understanding of
complex many-body systems~\cite{Cirac+Zoller:12}. In this context, trapped
ions~\cite{Leibfried+:03} constitute a versatile experimental platform on
which to explore a variety of models and phases~\cite{Blatt+Roos:12}. Owing
to the high degree of control over the internal degrees of freedom and the
single-ion addressability, trapped-ion systems allow both analog and digital
quantum simulations of quantum spin models~\cite%
{Porras+Cirac+Deng,Kim+Britton,Bermudez++:11}, models with bosonic degrees
of freedom~\cite{DengPorrasHaze}, and even of phenomena outside the realm of
low-energy physics~\cite{MilburnCasanovaHorstmannGerritsma}.

Electron-phonon (more generally, particle-phonon) coupling, a traditional
subject of solid-state physics, has only quite recently attracted interest
from the quantum-simulation community~\cite{PolaronSimulRefs}. The most
striking consequence of this coupling in semiconductors and insulators is
small-polaron formation: an excess electron (or a hole) can be severely
localized in the potential well that it creates by displacing the
surrounding atoms of the host crystal (``self-trapping'')~\cite%
{AlexandrovDevreese}. While the polaron concept was conceived by Landau and
Pekar in their studies of polar semiconductors~\cite{landaupekar}, evidence
has by now accumulated for small polarons in systems as diverse as amorphous
and organic semiconductors, manganites, undoped cuprates, superlattices of
graphene~\cite{PolaronMaterial}, and cold atomic gases~\cite{PolaronColdAtoms}. 
In particular, studies of small polarons in systems with short-range 
electron-phonon coupling are typically based on the Holstein molecular-crystal 
model~\cite{Holstein:59}. The static and dynamical properties of this model, 
describing purely local interaction of tightly-bound electrons with 
dispersionless (Einstein) phonons, have been extensively studied~\cite{AlexandrovDevreese}.

In this Letter, we present a scheme for simulating small-polaron formation
using a linear array of trapped ions subject to off-resonant standing light 
waves. We show that, if the relevant parameters are chosen appropriately, this
system is described by an effective model whose single-excitation sector
corresponds to the anti-adiabatic regime of the Holstein model (hopping
amplitude much smaller than the phonon frequency). Furthermore, we find that
with realistic values of experimental parameters (laser intensities,
detunings, trap frequencies, etc.) a regime is realized where the dressed
single-excitation ground state has polaronic character. Finally, we suggest
methods for measuring the quasiparticle spectral residue and 
the average number of phonons in the single-excitation ground state, 
quantities that provide typical signatures of small-polaron formation.
\begin{figure}[b!]
\includegraphics[bb=28 150 554 710,width=7.3cm,clip=]{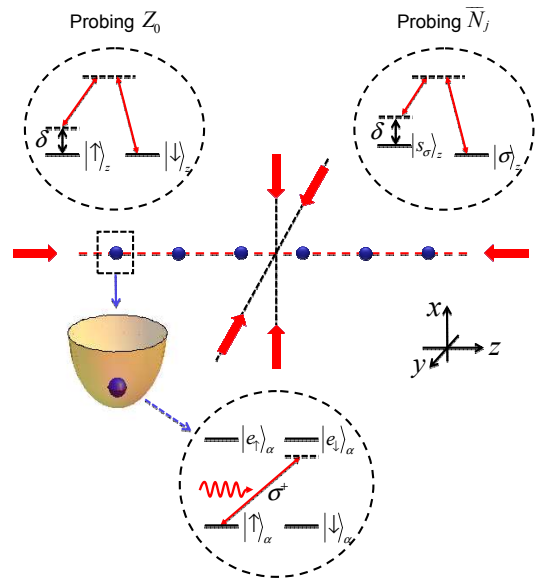}
\caption{\label{fig1}(Color online) System schematic: Trapped-ion chain subject to
three pairs of counterpropagating laser beams giving rise to three standing
waves. The standing wave with the polarization $\protect\sigma^{+}$ in the
direction $\protect\alpha$ induces the transition from $\left\vert\uparrow
\right\rangle_{\protect\alpha}$ to $\left\vert e_{\downarrow} \right
\rangle_{\protect\alpha}$, where $|\uparrow,\downarrow\rangle_{x}=
({|\uparrow\rangle_{z}\pm|\downarrow\rangle}_{z})/\sqrt{2}$ and
$|\uparrow,\downarrow\rangle_{y}=({|\uparrow\rangle_{z}\pm i|\downarrow\rangle}_{z})/\sqrt{2}$.}
\end{figure}

\textit{System and effective Hamiltonian}.-- We consider a
one-dimensional system of $N$ trapped ions with four internal states 
(the ground states $\left\vert\uparrow\right\rangle_{z}$ and $\left\vert\downarrow
\right\rangle_{z} $, as well as the excited states $\left\vert e_{\uparrow}\right\rangle_{z}$ 
and $\left\vert e_{\downarrow}\right\rangle_{z}$) and a possible physical
implementation as an array of microtraps (for an illustration, see Fig.~\ref{fig1}). 
We take the $z$-axis to be parallel to the direction of the ion chain
(longitudinal direction), while $x$ and $y$ will be the radial (transverse)
directions.

The total potential felt by an ion consists of the trapping-potential
contribution (with trapping frequencies $\omega_{\alpha}$, $\alpha=x,y,z$)
and the Coulomb repulsion between the ions. In the harmonic approximation,
the total vibrational Hamiltonian of the system reads (hereafter $\hbar=1$)
\begin{equation}
H_{\mathrm{ph}}=\sum_{i,\alpha }\frac{(p_{i}^{\alpha})^{2}}{2m}+\frac{m}{2}%
\sum_{i,j,\alpha}\mathcal{K}_{ij}^{\alpha}q_{i}^{\alpha}q_{j}^{\alpha} \:.
\end{equation}%
The displacements $q_{i}^{\alpha}$ of the $i$-th ion (mass $m$) from its
equilibrium position $z_{i}^{0}$ ($p_{i}^{\alpha }$ are the corresponding
momenta) are expressed through the creation and annihilation operators of
phonon modes with frequencies $\omega _{s,\alpha }$ as $q_{i}^{\alpha
}=\sum_{s}\mathcal{M}_{s,\alpha }^{i}(2m\omega _{s,\alpha
})^{-1/2}(a_{s,\alpha}+a_{s,\alpha }^{\dagger })$, such that $H_{\mathrm{ph}%
}=\sum_{s,\alpha}\omega _{s,\alpha}a_{s,\alpha }^{\dagger }a_{s,\alpha }$; $%
\mathcal{M}_{s,\alpha }^{i}$ is the $i$-th component of the normalized
eigenstate $\mathcal{M}_{s,\alpha }$ of the matrix $\mathcal{K}^{\alpha }$: $%
\sum_{i,j}\mathcal{M}_{s,\alpha }^{i}\mathcal{K}_{ij}^{\alpha }\mathcal{M}%
_{s^{\prime },\alpha }^{j}=\omega _{s,\alpha }^{2}\delta _{ss^{\prime }}$.
If the trapping potential in a transverse direction $\alpha$ dominates over
the Coulomb interaction [the stiff limit: $\beta_{\alpha}\equiv e^{2}/
(m\omega_{\alpha }^{2}d_{0}^{3})\ll 1$, where $d_{0}$ is the mean distance
between two adjacent ions], the vibrations in this direction are highly
localized in real space (weakly dispersive phonons)~\cite{Porras+Cirac+Deng}. 
On the other hand, by tuning $\omega_{z}$ one can reach both the
weakly-dispersive ($\beta _{z}\ll 1$) and the strongly-dispersive ($%
\beta_{z}\gg 1$) regime for longitudinal phonons.

As shown in Fig.~\ref{fig1}, three pairs of counterpropagating laser beams
with the polarization $\sigma^{+}$ form three standing waves, which induces
the transition from $\left\vert\uparrow\right\rangle _{\alpha }$ to $%
\left\vert e_{\downarrow}\right\rangle_{\alpha }$~\cite{Porras+Cirac+Deng}.
Under the rotating-wave approximation, the adiabatic elimination of the
higher energy states $\left\vert e_{\uparrow}\right\rangle_{z}$ and $%
\left\vert e_{\downarrow}\right\rangle_{z}$ leads to the effective
interaction%
\begin{equation}
H_{I}=\sum_{i,\alpha}\frac{G_{\alpha}^{2}}{2\Delta _{\alpha }}\cos
^{2}(k_{\alpha}q_{i}^{\alpha}+\phi_{\alpha})(1+\sigma_{i}^{\alpha })
\label{int_Ham}
\end{equation}%
between phonons and the pseudo-spins corresponding to the internal states $%
\left\vert \uparrow \right\rangle _{z}$ and $\left\vert \downarrow
\right\rangle _{z}$, where $G_{\alpha}$ is the Rabi frequency 
and $\Delta_{\alpha}\equiv\omega_{L,\alpha}-\omega_{0}$ 
the detuning between the energy-level spacing $\omega_{0}$ and the
laser frequency along the $\alpha$ direction; $\sigma_{i}^{\alpha}$ is
the Pauli matrix in the basis of $\left\vert \uparrow \right\rangle_{z}$ and 
$\left\vert\downarrow\right\rangle_{z}$. These pseudo-spins are
simultaneously acted upon by global magnetic fields, as described by the
Hamiltonian $H_{m}\equiv\sum_{i,\alpha}B_{\alpha}\sigma_{i}^{\alpha }$.

In the Lamb-Dicke limit, the ion-position dependent coupling constants can
be linearized around the equilibrium position, so that the interaction term
becomes $H_{I}=-\sum_{i,\alpha}F_{\alpha}q_{i}^{\alpha}(1+\sigma
_{i}^{\alpha})$. Here, the effective force $F_{\alpha}$ is given by 
$F_{\alpha}\sim G_{\alpha}^{2}k_{\alpha }/(2\Delta _{\alpha })$. For convenience, 
we focus on the parameters $F_{x}=F_{y}=F$, $\omega _{x}=\omega _{y}=\omega _{0}$, and $%
\beta_{x}=\beta_{y}=\beta$. Given that $F/(\omega_{0}\sqrt{2m\omega_{0}}%
)\equiv\eta\ll 1$, due to the large transverse-phonon frequencies, one can
perturbatively eliminate transverse phonons using a Fr\"{o}hlich-like
canonical transformation~\cite{Porras+Cirac+Deng}. In this manner, we obtain
the effective Hamiltonian%
\begin{eqnarray}
H_{\mathrm{eff}} &=&H_{L}+F_{z}\sum_{i}q_{i}^{z}(1+\sigma _{i}^{z})
\label{Heff} \\
&&+B_{z}\sum_{i}\sigma _{i}^{z}+\sum_{i<j,\alpha =x,y}\frac{2\beta \eta
^{2}\omega _{0}}{\left\vert i-j\right\vert ^{3}}\sigma _{i}^{\alpha }\sigma
_{j}^{\alpha },  \notag
\end{eqnarray}%
where the external magnetic fields $B_{\alpha }=F_{\alpha }^{2}/m\omega
_{\alpha }^{2}$ ($\alpha =x,y$) were chosen in order to cancel the effective
fields from the adiabatic elimination. The Hamiltonian%
\begin{equation}
H_{L}=\tilde{\omega}_{z}\sum_{i}b_{i}^{\dagger }b_{i}-\sum_{i\neq j}\frac{%
\tilde{\beta}_{z}\tilde{\omega}_{z}}{2\left\vert i-j\right\vert ^{3}}%
(b_{j}^{\dagger }b_{i}+\text{H.c.})  \label{H_L}
\end{equation}%
describes longitudinal phonons in terms of local-phonon operators, where
$q_{i}^{z}=(2m\tilde{\omega}_{z})^{-1/2}(b_{i}+b_{i}^{\dagger })$ and the
renormalized longitudinal-phonon frequency is given by $\tilde{\omega}_{z}=\omega
_{z}\sqrt{1+\sum_{j^{\prime}\neq i}(2\beta _{z}/\left\vert i-j^{\prime
}\right\vert^{3})}$. The form of the second term in Eq.~\eqref{H_L} results
from omitting the phonon-number nonconserving terms under the rotating-wave
approximation; the corresponding condition coincides with the stiff limit 
$\tilde{\beta}_{z}\equiv e^{2}/(m\tilde{\omega}_{z}^{2}d_{0}^{3})\ll 1$.

By switching to the spinless-fermion representation of the effective 
spin operators [the Jordan-Wigner transformation: $1+\sigma _{i}^{z}\rightarrow 2c_{i}^{\dagger }c_{i}$; $\sigma _{i}^{x}\sigma
_{j}^{x}+\sigma _{i}^{y}\sigma _{j}^{y}\rightarrow 2(c_{i}^{\dagger }c_{j}+%
\text{H.c.})$], the effective single-excitation Hamiltonian can be 
written as $H_{\mathrm{se}}=H_{\mathrm{e}}+H_{L}+H_{\mathrm{e-ph}}$, with 
\begin{eqnarray}
H_{\mathrm{e}} &=& \sum_{i<j}\frac{J}{\left\vert i-j\right\vert ^{3}}%
(c_{i}^{\dagger }c_{j}+\text{H.c.})+2B_{z}\sum_{i}c_{i}^{\dagger}c_{i}\:, \nonumber\\
H_{\mathrm{e-ph}} &=& g\tilde{\omega}%
_{z}\sum_{i}c_{i}^{\dagger }c_{i}(b_{i}+b_{i}^{\dagger}) \:.
\end{eqnarray}%
Here $J=4\beta \eta ^{2}\omega _{0}$ is the hopping amplitude, while 
$g=\sqrt{2}F_{z}/\sqrt{m\tilde{\omega}_{z}^{3}}$ is the dimensionless 
excitation-phonon (e-ph) coupling strength. This coupling, described 
by $H_{\mathrm{e-ph}}$, has Holstein-like (local) form. It is worthwhile 
to note that the above mapping of the effective spins to spinless fermions
is possible only in the one-dimensional case that we are addressing here.

\textit{Limits of validity and parameter range}.-- Let us summarize the
conditions of validity of the Hamiltonian $H_{\mathrm{se}}$ 
and specify in which parameter regime it reduces to the
Holstein model with dispersionless phonons.

Firstly, the use of perturbation theory for eliminating transverse phonons requires
that $\eta \ll 1$, which implies that $J\ll \omega _{0}$ must be fulfilled.
Secondly, while a weak phonon dispersion [here proportional to $\tilde{\beta}_{z}
\tilde{\omega}_{z}$; cf. Eq.~\eqref{H_L}] is inherent to the stiff limit,
for our purposes the dispersion should be negligible even compared to the
hopping amplitude $J$. From the condition $\tilde{\beta}_{z}\tilde{\omega}_{z}\ll J$
we readily obtain $\omega _{0}/\tilde{\omega}_{z}<4\eta ^{2}<1$. Thus our
system can simulate the Holstein model with $J\ll \omega _{0}<\tilde{\omega}%
_{z}$, i.e., in its anti-adiabatic regime. In other words, the effective
Hamiltonian of the system then reads $H=H_{\mathrm{e}}+H_{\mathrm{ph}}+H_{%
\mathrm{e-ph}}$, with $H_{\mathrm{ph}}=\tilde{\omega}_{z}\sum_{i}b_{i}^{%
\dagger }b_{i}$.

Unlike the genuine Holstein model, which only involves nearest-neighbor
hopping, the last effective Hamiltonian also contains long-range hopping
terms decaying with the characteristic inverse third power of the distance.
However, in the anti-adiabatic regime that we will be solely concerned
with these long-range terms are of minor importance.

We now discuss whether one can realize the strong-coupling regime of the
Holstein model, characterized by small-polaron formation. To achieve that,
a coupling constant $g$ of the order of unity is needed. We first express $g$
in terms of the relevant experimental parameters starting from the
expression $F_{z}\sim (G_{z}^{2}/\Delta _{z})\times (2\pi /\lambda )$, where
$\lambda$ is the laser wavelength, and $\Delta _{z}$ the detuning between the
laser frequency in the $z$ direction and the internal energy-level spacing. 
Since $g\tilde{\omega}_{z}=\sqrt{2}F_{z}/\sqrt{m\tilde{\omega}_{z}}$ and $a/d_{0}
\sim(\lambda\sqrt{2m\tilde{\omega}_{z}})^{-1}$, where $a$ is the size of the trap 
ground state (for concreteness, we assumed that $d_{0}=\lambda $, while in general 
$d_{0}$ is an integer multiple of $\lambda $), we readily obtain
\begin{equation}
g=\frac{4\pi}{\tilde{\omega}_{z}}\frac{G_{z}^{2}}
{\Delta_{z}}\times\frac{a}{d_{0}} \:.
\end{equation}%
Thus the value of $g$ can be varied by tuning the ratio $G_{z}^{2}/\Delta_{z}$,
for different values of the phonon frequency $\omega_{z}$.

We choose $\lambda=d_{0}=2\mu$m, and the typical value of the detuning $%
\Delta_{z}=1000$\:GHz. With Rabi frequencies $G_{z}=10-100$\:GHz, the ratio $%
G_{z}^{2}/\Delta_{z}$ is in the range $0.1-10$\:GHz. At the same time, 
the phonon frequency is typically $\omega _{z}/2\pi =1-20$\:MHz. For example, already for $%
G_{z}^{2}/\Delta_{z}=2$\:GHz and $\tilde{\omega}_{z}/2\pi =10$\:MHz (the ratio $%
a/d_{0}\approx 4.25\times 10^{-3}$) we obtain $g\approx 1.72$, a value of $g$
that belongs to the strong-coupling regime.

\textit{Polaron ground state}.-- To determine the ground state of the
system, we make use of the variational method. Variational approaches,
commonly employed in small-polaron studies, have been established to yield
results that agree very well with exact-diagonalization results on systems
of small size~\cite{PolaronVariational}.

The eigenstates of the total Hamiltonian are good quasi-momentum states,
i.e., simultaneous eigenstates of the total momentum operator 
$K=\sum_{k}k\:c_{k}^{\dagger }c_{k}+\sum_{q}q\:b_{q}^{\dagger}b_{q}$ (with
eigenvalues denoted by $\kappa $ in what follows). Therefore, variational
states must be Bloch-like linear combinations ${|{\psi }_{\kappa }\rangle }%
=N^{-1/2}\sum_{n}e^{i\kappa n}{|{\psi }_{\kappa }(n)\rangle }$ of
Wannier-like functions ${|{\psi}_{\kappa}(n)\rangle}$. In particular, for
the Toyozawa Ansatz states~\cite{Toyozawa:61}
\begin{equation}
{|{\psi }_{\kappa }(n)\rangle }=\sum_{m=-N/2}^{N/2-1}{\Phi }_{\kappa
}(m)e^{i\kappa m}c_{n+m}^{\dg}{|0\rangle }_{\mathrm{e}}\otimes {|\xi
_{\kappa }(n)\rangle }_{\mathrm{ph}},
\end{equation}%
with ${|\xi _{\kappa }(n)\rangle }_{\mathrm{ph}}\equiv \prod_{l}\exp \big(%
v_{l}^{\kappa }b_{n+l}^{\dagger }-v_{l}^{\kappa \ast }b_{n+l}\big)|0\rangle
_{\mathrm{ph}}$ being a direct product of phonon coherent states at sites $%
n+l$ ($l=-N/2,\ldots ,N/2$). The parameters $\{{\Phi}_{\kappa
}(m),v_{l}^{\kappa }\}$, in total $2N$ of them, are complex valued. The
normalization is given by ${\langle {\psi }_{\kappa }|{\psi }_{\kappa
}\rangle }=\sum_{m,m^{\prime }}{\Phi }_{\kappa }^{\ast }(m^{\prime }){\Phi }%
_{\kappa}(m)\;S_{m^{\prime}-m}^{\kappa}$, where $S_{n-n^{\prime
}}^{\kappa}\equiv\exp\big[-\frac{1}{2}\sum_{p}\big(|v_{p-n^{\prime
}}^{\kappa}|^{2}+|v_{p-n}^{\kappa }|^{2}-2v_{p-n^{\prime }}^{\kappa \;\ast
}v_{p-n}^{\kappa }\big)\big]$.

In the limit $\tilde{\omega}_{z}/J\rightarrow\infty$ the Toyozawa Ansatz
yields (for $\kappa=0$) the exact ground state of the Holstein model, given
by $a^\dg_{n_{0}}{\vert 0 \rangle}_{\mathrm{e}}\otimes{\vert v^{\kappa
=0}_{n_{0}}=g\rangle}_{\mathrm{ph}}$ (where $v_l^{\kappa=0}$ at sites 
$l\neq n_{0}$ are equal to zero)~\cite{PolaronVariational,Toyozawa:61}. The last statement holds true
not only for the genuine Holstein model with nearest-neighbor hopping $J$,
but also for our model with long-range hoppings, as the latter have
amplitudes proportional to $J$. This makes the Toyozawa Ansatz particularly
appropriate for the problem at hand.

The ground-state energy $E_{\text{GS}}$ is obtained by minimizing the
expectation value ${\langle{\psi}_{\kappa=0}\vert}H{\vert{\psi}%
_{\kappa=0}\rangle}$ with respect to $4N$ variational parameters, the
modules and phases of the complex parameters ${\Phi}_{\kappa=0}(m)$ and $%
v^{\kappa=0}_{l}$. While there is no principal obstacle to treating even
larger system, in the following we present results obtained for $N=32$.

\begin{figure}[t!]
\includegraphics[width=7.2cm]{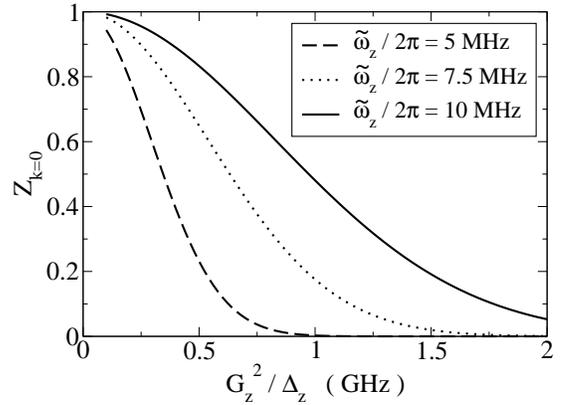}
%%[trim=40mm 1mm 40mm 1mm,clip,width=1.75cm]
\caption{Calculated quasiparticle residue at $k=0$ for different values of
the renormalized longitudinal-phonon frequency $\tilde{\protect\omega}_{z}$.
All three curves correspond to the hopping amplitude $J=25$\:KHz.}
\label{Z_k=0}
\end{figure}
\textit{Characterization of the polaron crossover}.-- The quasiparticle residue
at quasi-momentum $k$ is defined as $Z_{k}\equiv|\langle\Psi_{k}|\psi_{k}%
\rangle|^{2}/\langle\psi_{k}|\psi_{k}\rangle$, the overlap squared of the
bare-excitation Bloch state $|\Psi_{k}\rangle\equiv
c^{\dagger}_{k}|0\rangle=N^{-1/2}\sum_{n}e^{ikn}
c^{\dagger}_{n}|0\rangle$ and the corresponding (dressed-excitation) Bloch
state $|\psi_{k}\rangle$ of the coupled excitation-phonon system. For the
Toyozawa Ansatz state, we find
\begin{equation}  \label{defZ}
Z_{k}=\frac{\displaystyle\Big|\sum_{m}{\Phi}_{k}(m)\Big|^{2}} {\displaystyle%
\sum_{m,m^{\prime}}{\Phi}_{k}^*(m^{\prime}){\Phi}_{k} (m)\;
S^{k}_{m^{\prime}-m}}\:\prod_{l}e^{-|v^{k}_{l}|^{2}}\:.
\end{equation}
The quasiparticle residue $Z_{k=0}$, when evaluated for optimal values of
the variational parameters, characterizes the ground state of the system. The
change of this quantity from unity (non-interacting system) to values very
close to zero (Fig.~\ref{Z_k=0}) illustrates the crossover from a quasi-free
excitation to a small polaron.

Another quantity characterizing the polaron crossover is the average number of
phonons in the ground state
\begin{multline}  \label{phave}
\bar{N}_{\text{ph}}\equiv{\langle{\psi}_{\kappa=0}\vert}\:\sum_{i}
b^{\dagger}_{i}b_{i}\:{\vert{\psi}_{\kappa=0}\rangle} \\
=\sum_{m,m^{\prime},p}{\Phi}_{\kappa=0}^{*}(m^{\prime}) {\Phi}_{\kappa=0}
(m)S^{\kappa=0}_{m^{\prime}-m}v^{\kappa=0*}_{p+m^{\prime}}
v^{\kappa=0}_{p+m}\:.
\end{multline}

\begin{figure}[t!]
\includegraphics[width=7.2cm]{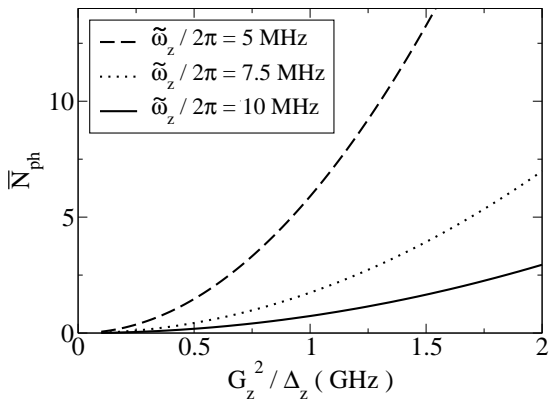}
%%[trim=40mm 1mm 40mm 1mm,clip,width=1.75cm]
\caption{Average number of phonons in the single-excitation ground state.
All three curves correspond to $J=25$\:KHz.}
\label{NphFig}
\end{figure}
\noindent As illustrated in Fig.~\ref{NphFig}, this quantity varies from $%
\bar{N}_{\text{ph}}\approx 0$ (quasi-free excitation) to values $\bar{N}_{%
\text{ph}}\gtrsim 3$ (small polarons) with increasing coupling strength.

Our analysis indeed shows that the effect of the fast-decaying long-range hopping
on the quantities characterizing the polaron crossover is very small in the
anti-adiabatic regime. This is consistent with a recent study of the Holstein 
model with next-nearest-neighbor hopping which established that such hopping 
is significant only in the opposite, adiabatic regime~\cite{Chakraborty+:11}.

\textit{Experimental validation}.-- We propose two experimental methods, as
illustrated in Fig.~\ref{fig1}, to detect the quasiparticle residue $Z_{k=0}$ and
the average number of phonons $\bar{N}_{\text{ph}}$ in the single-excitation 
ground state. 

To probe $Z_{k=0}$, we first prepare the system in the ground state
(without a polaronic excitation), namely, the vacuum state of the longitudinal phonons and
the internal states $\left\vert\downarrow\right\rangle$ of all the ions. Then,
all the laser beams are turned on. At the same time, an additional Raman
laser with the detuning $\delta$ is used for generating an excitation (polaron).
The latter is described by the Hamiltonian $H_{\text{detect}}=\xi
N^{-1/2}\sum_{i}(\sigma _{i}^{+}e^{-i\delta t}+$H.c.$)$, where $\xi$ is the
effective Rabi frequency. After the evolution
time $T$, we switch off the laser beams and measure $O=N^{-1/2}\sum_{i}\sigma
_{i}^{x}$~\cite{Leibfried+:03}. By repeating the measurement one can
determine the mean-value $\left\langle O\right\rangle$. For short times $%
T\ll 1/\xi$, near the resonance frequency $\delta\sim E_{\mathrm{GS}}$ 
of the ground state in the single-polaron space, first-order perturbation 
theory yields
\begin{equation}
\left\langle O\right\rangle\sim 2\xi Z_{k=0}\frac{\sin(\delta T)
-\sin(E_{\mathrm{GS}}T)}{\delta-E_{\mathrm{GS}}}\:.
\end{equation}%
Thus measurements of $\left\langle O\right\rangle$ provide a direct access 
to the quasiparticle residue $Z_{k=0}$, the quantity whose strongly reduced 
values indicate the presence of small polarons.

For detecting the mean phonon number $\bar{N}_{\text{ph}}$, we first prepare 
the system in the single-polaron ground state 
$\left\vert E_{\mathrm{GS}}\right\rangle\equiv\vert{\psi}_{\kappa=0}\rangle$ 
and switch off all laser beams. Then, we detect the internal
state of the ion on the site $j$. If the internal state is $\left\vert
\downarrow \right\rangle _{j}$, the Raman laser is turned on to induce the
transition between $\left\vert\downarrow\right\rangle _{j}$ to another
state $\left\vert s_{\downarrow }\right\rangle _{j}$; if the internal state
is $\left\vert \uparrow \right\rangle _{j}$, the Raman laser is turned on to
induce the transition between $\left\vert \uparrow \right\rangle _{j}$ to
the state $\left\vert s_{\uparrow}\right\rangle _{j}$, as shown in Fig.~%
\ref{fig1}. Here, each Raman laser is turned to the lower sideband detuning $%
\delta =\omega _{z}$. Thus, in the interaction picture the time evolution of
the system is described by the Hamiltonian $H_{\sigma=\uparrow,\downarrow
}=\xi(b_{j}\left\vert s_{\sigma }\right\rangle_{j}\left\langle \sigma
\right\vert +$H.c.$)$. After the evolution time $T$, we detect the
probability $p_{\sigma,j}$ of the transition to $\left\vert s_{\sigma }\right\rangle
_{j}$. For the short-time evolution, first-order perturbation theory
leads to $p_{\sigma,j}=\xi^{2}T^{2}\left\langle E_{\mathrm{GS}}\right\vert
b_{j}^{\dagger }b_{j}P^{(j)}_{\sigma }\left\vert E_{\mathrm{GS}}\right\rangle $,
where $P^{(j)}_{\sigma}=\left\vert\sigma\right\rangle_{j}\left\langle \sigma
\right\vert$ is the projection operator. Finally, the average number of
phonons at site $j$ is $\bar{N}_{j}=\sum_{\sigma }p_{\sigma,j}/(\xi
^{2}T^{2}) $. Therefore, measurements of the transition probabilities $p_{\sigma,j}$ 
($j=1,\ldots,N$) allow us to extract the average number of phonons at each site (microtrap).
Large values of $\bar{N}_{\text{ph}}=\sum_{j}\bar{N}_{j}$ represent a signature of small polarons.

\textit{Multiple-excitation regime}.-- Apart from simulating the conventional
single-polaron problem, our envisioned system in the multiple-excitation regime 
should allow one to study the phenomenon of density-driven destabilization 
of small polarons~\cite{EminPolaronDestabil} in a particularly
clean manner. Namely, the spinless-fermion character of excitations this one-dimensional 
system precludes phenomena typical of electron-phonon interaction (e.g., formation of intrasite 
bipolarons~\cite{AlexandrovDevreese}) or its interplay with Hubbard-type on-site electron 
repulsion (spin- or charge density waves) in many-electron solid state systems.
 
Quite generally, the driving force for excitations to undergo self-trapping 
decreases with their density due to destructive interference of phonon
displacements originating from adjacent small polarons. This in turn
effectively lowers the binding energy of those self-trapped excitations. 
Simple qualitative arguments show that in a disorder-free system all 
excitations remain quasifree above the critical density 
(excitation number per site)~\cite{EminPolaronDestabil}  
\begin{equation}
n_c \approx 1-\frac{W}{2E_b} \:, \label{crit_density}
\end{equation}
where $E_b\equiv g^{2}\omega$ (with $\omega$ being the relevant
phonon energy) is the small-polaron binding energy 
in the low-density limit, and $W$ the bare-excitation bandwidth. Note
that $E_b \ge W/2$ is the necessary condition for the existence of 
a single small polaron~\cite{AlexandrovDevreese}.

In our system $\omega=\tilde{\omega}_{z}$, with $W\approx 4J$ and $J\ll\tilde{\omega}_{z}$, 
thus implying that $W\ll E_b$. Therefore, Eq.~\eqref{crit_density} yields a 
critical density slightly below the maximal density of one particle per site. 
Our trapped-ion system, where the excitation density is an accessible experimental
knob and polaronic behavior (or lack thereof) can be detected as explained above, 
thus allows for a quantitative study of this conceptually important problem of 
small polarons in the high-density limit.

\textit{Conclusions}.-- To conclude, we have suggested a scheme for the quantum
simulation of small-polaron physics with trapped ions. We have shown that with
realistic values of the relevant experimental parameters one can reach
the strong-coupling regime of the Holstein model, characterized by
small-polaron formation. Besides, we have anticipated that the trapped-ion scheme 
proposed here opens up possibilities to go beyond the single-excitation regime 
and study more complex phenomena, e.g., behavior of small polarons in 
the high-density limit. Furthermore, with the advent of surface 
traps~\cite{Leibfried+:03,SurfaceTraps} it should also be possible 
to engineer other kinds of interactions. Experiments with trapped ions may 
allow us to better understand polaron physics in these circumstances, and 
to develop new theoretical tools to describe the resulting phenomena.

\begin{acknowledgments}
V.M.S. and C.B. were supported by the Swiss NSF, the NCCR QSIT, and the NCCR
Nanoscience. T. Shi and J. I. Cirac acknowledge financial support from the
EU project AQUTE and DFG (SFB631).
\end{acknowledgments}

\end{document}